%% file: main.tex
\documentclass[aps,prb,reprint,
               superscriptaddress,
               floatfix,
               amsmath,amssymb,
               longbibliography]{revtex4-2}
\AtBeginDocument{\hypersetup{colorlinks=true,linkcolor=black,citecolor=black,urlcolor=black}}
\usepackage{graphicx}
\usepackage{dcolumn}
\usepackage{bm}
\usepackage{xcolor}
\usepackage{pifont}
\usepackage{booktabs}
\usepackage{multirow}
\usepackage{array}
\usepackage{float}
\usepackage{orcidlink}

\begin{document}
\preprint{APS/123-QED}

\title{Electronic correlations shape the low-energy optical response of the \\ kagome antiferromagnets Mn$_3$Sn and Mn$_3$Ge}

\author{R.~Mathew Roy\orcidlink{0009-0003-2842-9044}}
   \thanks{Contact author: renjith.mathew-roy@pi1.uni-stuttgart.de}
\affiliation{1.~Physikalisches Institut, Universit\"at Stuttgart, Pfaffenwaldring 57,
70569 Stuttgart, Germany}

\author{Bo Tai}
\affiliation{Max Planck Institute for Chemical Physics of Solids, 01187 Dresden, Germany}

\author{Maxim~Wenzel\orcidlink{0009-0000-8813-2080}}
\affiliation{1.~Physikalisches Institut, Universit\"at Stuttgart, Pfaffenwaldring 57,
70569 Stuttgart, Germany}

\author{Achyut~Tiwari\orcidlink{0009-0007-2892-5012}}
\affiliation{1.~Physikalisches Institut, Universit\"at Stuttgart, Pfaffenwaldring 57,
70569 Stuttgart, Germany}

\author{Mykhaylo~Ozerov\orcidlink{0000-0002-5470-1158}}
\affiliation{National High Magnetic Field Laboratory, Florida State University, 1800 E Paul Dirac Dr, Tallahassee, Florida 32310, USA}

\author{Chandra~Shekhar\orcidlink{0000-0002-3330-0400}}
\affiliation{Max Planck Institute for Chemical Physics of Solids, 01187 Dresden, Germany}

\author{Claudia~Felser\orcidlink{0000-0002-8200-2063}}
\affiliation{Max Planck Institute for Chemical Physics of Solids, 01187 Dresden, Germany}
 
\author{Artem V. Pronin\orcidlink{0000-0003-2585-7476}}
\affiliation{1.~Physikalisches Institut, Universit\"at Stuttgart, Pfaffenwaldring 57,
70569 Stuttgart, Germany}

\author{Xiaolong~Feng}
\affiliation{Max Planck Institute for Chemical Physics of Solids, 01187 Dresden, Germany}

\author{Martin~Dressel\orcidlink{0000-0003-1907-052X}}
\affiliation{1.~Physikalisches Institut, Universit\"at Stuttgart, Pfaffenwaldring 57,
70569 Stuttgart, Germany}

\begin{abstract}

Using optical spectroscopy and density functional theory, we provide evidence that electron-electron interactions strongly affect the properties of the kagome antiferromagnet Mn$_3$Sn, since its low-energy optical interband transitions arise exclusively from the correlation-modified electronic band structure. Mn$_3$Sn possesses an optical effective mass about three times larger than that of its isostructural analog Mn$_3$Ge. The DFT+$U$ treatment in the Lichtenstein formulation, with an on-site Coulomb repulsion $U = 4$~eV and Hund's coupling $J = 0.25$~eV, accounts for the optical transitions in Mn$_3$Sn, whereas those in Mn$_3$Ge are already reproduced without Hubbard corrections ($U, J = 0$). The near-isotropic electronic structure of Mn$_3$Sn reveals a three-dimensional metal whose electronic response is shaped prominently by the Mn $d$-states. The apparent linear regime in $\sigma_1(\omega)$ observed in Mn$_3$Sn and Mn$_3$Ge arises from several overlapping interband transitions and therefore should not be interpreted as the optical signature of the Weyl cones. The optical response does not alter with magnetic field up to 17 T, consistent with strong free-carrier screening. Our findings establish a comprehensive-correlated picture of Mn$_3$Sn and Mn$_3$Ge that offers a template for other correlated topological metals.

\end{abstract}

\date{\today}
\maketitle

%-------------------------------------------------------------------
\section{Introduction}

Antiferromagnetic spintronics is emerging as a route toward terahertz-fast, magnetically robust memory devices, owing to the vanishingly small stray fields of antiferromagnets~\cite{jungwirth2016antiferromagnetic,baltz2018antiferromagnetic,lee20191gbit,aggarwal2019demonstration}. From a fundamental standpoint, magnetic topological metals promise to combine this magnetic robustness with topologically protected charge and spin transport. When magnetic order coexists with nontrivial band topology, the Berry curvature of the occupied bands drives transverse charge, heat, and spin currents that persist without net magnetization and are readily tunable through the spin order~\cite{nagaosa2010anomalous,vsmejkal2018topological}, manifesting as measurable electrical and optical responses. Berry curvature effects are, however, functionals of the underlying electronic structure, which may be significantly modified by sizeable on-site Coulomb repulsion~\cite{basovelectrodynamics, xu2020electronic, wenzel2022effect}.

The noncollinear antiferromagnets Mn$_3$Sn and Mn$_3$Ge are canonical realizations of this interplay: Their inverse-triangular spin order carries a ferroic cluster-magnetic-octupole moment that breaks time-reversal symmetry and, through spin-orbit coupling, permits nonzero intrinsic transverse responses. This, in turn, has enabled demonstrations of functional devices~\cite{yoon2025electrical,zheng2024effective}. First-principles calculations have predicted Weyl nodes near the Fermi level~\cite{yang2017topological}, while experiments have reported large anomalous Hall~\cite{nakatsuji2015large,nayak2016large,kiyohara2016giant}, anomalous Nernst~\cite{ikhlas2017large,li2017anomalous}, and magneto-optical Kerr effects~\cite{higo2018large}. Weyl nodes may enhance the Berry curvature, but these measurements do not by themselves uniquely establish Weyl topology. All of these observables are governed by interband Berry curvature summed over the Brillouin zone, and therefore inherit the details of the underlying band structure. Understanding the interplay between electronic correlations, magnetic ordering, and topological effects consequently remains the central challenge.

%---------------------------------------------------
%Fig1

\begin{figure*}
    \centering
    \includegraphics[width=1.0\textwidth]{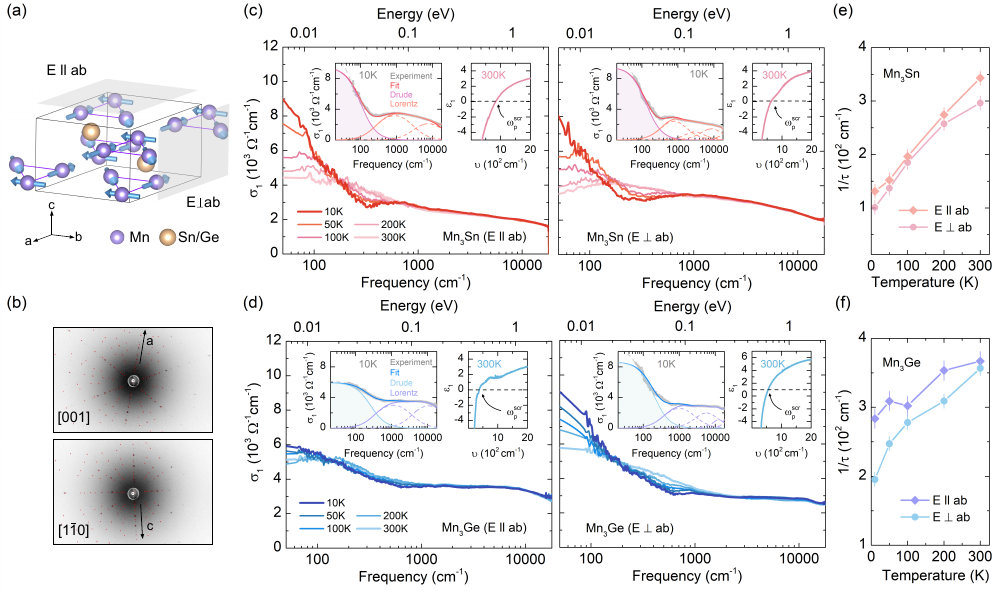}
    \caption{(a) Crystal structure of Mn$_{3}X$ ($X =$ Sn, Ge), with the two measurement geometries indicated: $\mathrm{E}\parallel ab$ (in-plane) and $\mathrm{E}\perp ab$ (out-of-plane). (b) x-ray diffraction patterns of Mn$_{3}$Ge measured on the (001) and (1$\bar{1}$0) oriented surfaces. (c)~Temperature dependence of the real part of the optical conductivity $\sigma_{1}(\omega)$ of Mn$_{3}$Sn for $\mathrm{E}\parallel ab$ (left panel) and $\mathrm{E}\perp ab$ (right panel). (d) Same as (c) for Mn$_{3}$Ge. The left inset of each panel in (c) and (d) shows the Drude--Lorentz decomposition of $\sigma_1(\omega)$ at 10~K, with the measured spectrum, the total fit, and the individual Drude and Lorentz contributions. The right inset shows the real part of the dielectric function $\varepsilon_1(\omega)$ at 300~K; its zero crossing defines the screened plasma frequency $\omega_p^{\mathrm{scr}}$. (e) Temperature dependence of the scattering rate $1/\tau$ of Mn$_{3}$Sn along the in-plane and out-of-plane directions. (f) Same as (e) for Mn$_{3}$Ge.}
  \label{opticalcond}
\end{figure*}
%---------------------------------------------------

Hexagonal Mn$_{3}$Sn and Mn$_{3}$Ge both crystallize in the Ni$_{3}$Sn-type structure (space group $P6_3/mmc$, No.~194), in which the Mn ions form kagome layers in the $(001)$ plane stacked along the $c$ axis, with Sn or Ge occupying the layer-centered sites~\cite{tomiyoshi1982magnetic,brown1990determination,nagamiya1982triangular,rai2022unconventional}, as displayed in Fig.~\ref{opticalcond}(a) together with the x-ray diffraction patterns of Mn$_{3}$Ge in Fig.~\ref{opticalcond}(b). Geometrical frustration of the kagome plaquettes, balanced against Dzyaloshinskii--Moriya interactions, stabilizes a noncollinear inverse-triangular $120^{\circ}$ antiferromagnetic texture below $T_{\mathrm{N}}\approx 420~\mathrm{K}$ in Mn$_{3}$Sn and $365~\mathrm{K}$ in Mn$_{3}$Ge~\cite{brown1990determination,nagamiya1982triangular}. In Mn$_{3}$Sn, another first-order transition near $270~\mathrm{K}$ drives the system into an incommensurate helical phase that quenches the anomalous Hall effect~\cite{sung2018magnetic,park2018magnetic}. Moreover, the intralayer connectivity and out-of-plane stacking produce a strong magnetocrystalline anisotropy~\cite{duan2015magnetic,rai2022unconventional,gao2024anisotropic,gao2025anisotropic}.

Angle-resolved photoemission spectroscopy (ARPES) on Mn$_3$Sn reports strongly smeared, poorly resolved bands together with a band renormalization factor of ${\sim}5$, which is widely regarded as an indication of pronounced electronic correlations~\cite{kuroda2017evidence}.
The isostructural Mn$_3$Ge, by contrast, shows a renormalization of only ${\sim}1.18$ and has been characterized as weakly correlated~\cite{changdar2024weak}. Subsequent density functional theory combined with dynamical mean-field theory (DFT+DMFT) calculations for Mn$_3$Sn showed that including correlations modifies the bands, shifts the Weyl points, and pushes weakly dispersive (flat) bands toward the Fermi level, leading to band-specific renormalization effects~\cite{yu2022correlated}. With DFT+DMFT, the renormalization is not uniform but band specific. More recently, however, a nearly universal band renormalization factor of $Z^{-1} \approx 1.6$ has been reported, attributing the band and magnetic-moment evolution to Hund's coupling ($J$) rather than only Coulomb repulsion $U$~\cite{cao2026correlated}. Settling these competing estimates therefore demands a bulk, momentum-integrated probe rather than relying on comparisons with momentum-resolved ARPES alone.

The effect of correlations can significantly shapes the optical response, a connection well established for heavy-fermion metals and high-$T_c$ superconductors~\cite{basovelectrodynamics}. We use bulk-sensitive optical spectroscopy to disentangle the free-carrier and interband contributions and to assess the role of the Hubbard $U$ and Hund's coupling $J$ by comparing the measured interband features with DFT. No broadband single-crystal optical study of hexagonal Mn$_3$Ge has been reported, and the only such study of Mn$_3$Sn~\cite{Cao2023optical} was interpreted without reference to band structure or correlation. We reveal that the experimental low-energy optical conductivity of Mn$_3$Ge is well reproduced by standard DFT calculations ($U=J=$0), while the optical response of Mn$_3$Sn is significantly reshaped by the presence of correlations ($U = 4$~eV, $J = 0.25$~eV), that accounts for the experimental low-energy optical conductivity. Both compounds display Fermi-liquid-like behavior across a broad Drude regime. Together, our results confront electronic structures (with and without correlations) and reveal the natural correlated states of Mn$_3$Sn and Mn$_3$Ge.

%Fig2

\begin{figure}
    \centering
    \includegraphics[width=0.45\textwidth]{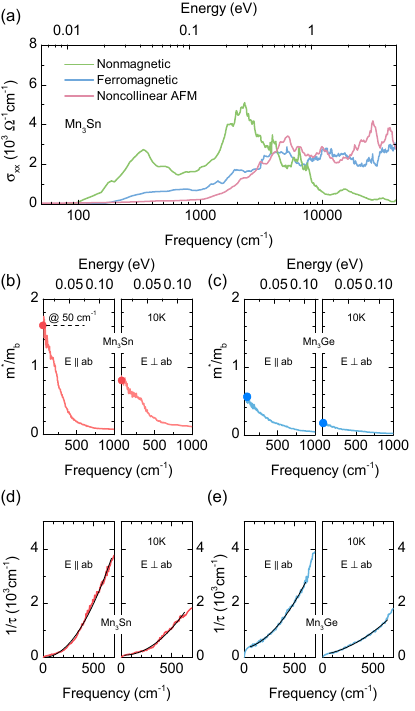}
    \caption{(a) The calculated optical conductivities ($\sigma_{xx}$) of Mn$_3$Sn for nonmagnetic, ferromagnetic, and noncollinear order of the Mn moments. (b) The frequency dependence of the effective mass of Mn$_{3}$Sn for the in-plane and out-of-plane directions. (c) Same as (b), but for Mn$_{3}$Ge. (d) The frequency dependence of the scattering rate for Mn$_{3}$Sn; the black line is a quadratic polynomial fit. (e) Same as (d), but for Mn$_{3}$Ge.}
  \label{effcetivemass}
\end{figure}

%-------------------------------------------------------------------

%-------------------------------------------------------------------
\section{Methodology}

\subsection{Experimental details}

Mn$_3$Sn and Mn$_3$Ge single crystals were prepared by the Bridgman--Stockbarger technique as described in the Ref.~\cite{nayak2016large}. The starting compositions were Mn:Ge $=$ 3.08:0.92 for Mn$_3$Ge and Mn:Sn $=$ 3:1 for Mn$_3$Sn. For both compounds, the melt was cooled from 1100~$^\circ$C at a rate of 2~$^\circ$C/h to promote single crystal growth.

The infrared reflectivity of freshly polished crystals was measured in the frequency range from 80 to 18\,000~cm$^{-1}$ ($10~\mathrm{meV}$ to $2.2~\mathrm{eV}$), as a function of temperature from $T=300~\mathrm{K}$ down to $10~\mathrm{K}$. Measurements were performed for two incident light polarizations, $\mathbf{E} \parallel ab$ (parallel to the kagome plane) and $\mathbf{E} \perp ab$ (perpendicular to the kagome plane), corresponding to the $(001)$ and $(1\bar{1}0)$ surfaces and denoted ``in-plane'' and ``out-of-plane,'' respectively. Within the $(1\bar{1}0)$ surface the measured reflectivity showed no significant dependence on the polarizer orientation relative to the $c$ axis over the full spectral range. The real parts of the optical conductivity $\sigma_1(\omega)$ and of the dielectric function $\varepsilon_1(\omega)$ were obtained from the measured reflectivity via a Kramers--Kronig analysis (shown in Fig.~S1 of the Supplemental Material (SM)~\cite{SM}). 

Additionally, room-temperature ellipsometric measurements were performed on a dual-rotating-compensator ellipsometer (RC2, J.~A. Woollam Co., Inc.) over the energy range $0.73$--$5~\mathrm{eV}$. Since the intraband and interband contributions are not separated by a well-defined energy scale, we applied both a Drude--Lorentz decomposition and an extended Drude analysis; the latter yields $m^{*}/m_b(\omega)$ and $1/\tau(\omega)$ directly from the complex conductivity.

\subsection{Computational details} 

The electronic structures of Mn$_3$Sn and Mn$_3$Ge were calculated within density functional theory using the projector-augmented-wave method as implemented in VASP~\cite{kresse1996efficient,kresse1999paw}. Exchange and
correlation were described using the Perdew--Burke--Ernzerhof generalized-gradient approximation~\cite{perdew1996pbe}. Spin-orbit coupling was included self-consistently in noncollinear calculations of the coplanar $120^{\circ}$ Mn spin structure. Correlation effects in the Mn-$3d$ subspace were treated using the rotationally invariant Lichtenstein DFT+$U$ functional~\cite{liechtenstein1995density}. The correlated band structures were calculated with $U=4$~eV and $J=0.25$~eV and compared with calculations without a Hubbard correction.

A plane-wave cutoff of 450~eV and a $9\times9\times7$
$\Gamma$-centered $k$-point mesh were used for the self-consistent calculations. The total energy was converged to $10^{-8}$~eV, and Gaussian smearing of 0.01~eV was applied. Starting from the experimental structures~\cite{brown1990determination,kiyohara2016giant,rai2022unconventional}, the relaxed lattice parameters were $a = 5.665$~\AA{} and $c = 4.531$~\AA{} for Mn$_3$Sn and $a = 5.302$~\AA{} and $c = 4.289$~\AA{} for Mn$_3$Ge. For Mn$_3$Sn/Mn$_3$Ge, the interband contributions to $\operatorname{Re}\sigma_{xx}(\omega)$ and $\operatorname{Re}\sigma_{zz}(\omega)$ were evaluated from a spinor Wannier Hamiltonian constructed with Wannier90~\cite{pizzi2020wannier90}, followed by evaluation of the complex conductivity tensor. The optical response was evaluated at 10~K using a Lorentzian broadening of 5~meV and an $8\times8\times13$ $k$-point mesh.  The photon-energy range extended from 0.01 to 3.00~eV.  Band-pair contributions were evaluated on a $3\times3\times5$ mesh.

%---------------------------------

\begin{table*}[t]
\caption{Experimental and DFT plasma frequencies $\omega_p$ (in eV) for Mn$_3$Sn and Mn$_3$Ge along the $xx$ and $zz$ polarization directions. The experimental values $\omega_p^{\rm exp}$ have an uncertainty of $\pm 0.~3$~eV. DFT values are given for $U=0$ and $U=4$\,eV. The ratio $(\omega_p^{\mathrm{exp}})^2/(\omega_p^{\mathrm{DFT}})^2$ and $(\omega_p^{\mathrm{exp}})^2/(\omega_p^{\mathrm{DFT+\it{U}}})^2$ quantifies the mass renormalization without and with the Hubbard $U$, respectively.}
\label{tab:plasma}
\begin{ruledtabular}
\begin{tabular}{lccccc}
 & $\omega_p^{\mathrm{exp}}$
 & $\omega_p^{\mathrm{DFT}}$
 & $\omega_p^{\mathrm{DFT+\it{U}}}$
 & $(\omega_p^{\mathrm{exp}})^2/(\omega_p^{\mathrm{DFT}})^2$
 & $(\omega_p^{\mathrm{exp}})^2/(\omega_p^{\mathrm{DFT+\it{U}}})^2$  \\
\hline
Mn$_3$Sn ($xx$) & 3.40 & 1.58 & 3.80 & 4.63 & 0.80 \\
Mn$_3$Sn ($zz$) & 2.80 & 1.57 & 4.05 & 3.18 & 0.47 \\
Mn$_3$Ge ($xx$) & 1.30 & 1.53 & 2.37 & 0.72 & 0.30 \\
Mn$_3$Ge ($zz$) & 1.80 & 2.48 & 2.96 & 0.52 & 0.37 \\
\end{tabular}
\end{ruledtabular}
\end{table*}

%-------------------

\section{Results and Discussion}

\subsection{Charge dynamics, magnetic configuration,\\ and field response of Mn$_3$Sn}

The temperature evolution of the real part of the optical conductivity of Mn$_3$Sn is shown in Fig.~\ref{opticalcond}(c) for the in-plane ($\mathbf{E} \parallel ab$) and out-of-plane ($\mathbf{E} \perp ab$) directions. The high-energy part of $\sigma_1(\omega)$ ($\omega/(2\pi c)>1000~\mathrm{cm}^{-1}$), which arises from interband transitions, shows only a weak temperature dependence, pointing to the absence of any significant band modifications with temperature. In particular, the interband features of Mn$_3$Sn remain intact across the helical ordering transition near $270~\mathrm{K}$~\cite{cable1993neutron,chen2024intertwined}, as previously noted by Cao \textit{et al.}~\cite{Cao2023optical}. Moreover, these interband transitions are strongly broadened by the large scattering rate, resembling Heusler analogs~\cite{shreder2024optical,Cao2023optical}. Also, the interband-response anisotropy (in-plane vs out-of-plane) is weak, indicating that from an optical standpoint Mn$_3$Sn behaves as an essentially three-dimensional bulk kagome system.

The various electronic contributions of Mn$_3$Sn are well captured by a Drude--Lorentz analysis. The 10~K spectra, shown in the inset of Fig.~\ref{opticalcond}(c), are described, to a good approximation, by a single Drude term, while the multiple Lorentzians account for the high-energy interband transitions. The momentum-relaxation rate $1/\tau$ of the Drude component, displayed in Fig.~\ref{opticalcond}(e), remains nearly identical for the in-plane and out-of-plane directions, and its temperature dependence follows the electrical transport \cite{tomiyoshi1987electrical}, with $1/\tau$ at 10~K reduced by $\sim$70\% relative to its 300~K value. The free-carrier scattering response is thus essentially isotropic. In contrast, the anomalous responses reported for Mn$_3$Sn are strongly direction-selective~\cite {nakatsuji2015large,chen2021anomalous}. The absence of a comparable anisotropy in the longitudinal charge channel suggests that this direction dependence of anomalous responses is encoded in the noncollinear Mn-moment configuration, acting through spin-orbit coupling, rather than in the itinerant charge channels \cite{suzuki2017cluster}.

The DFT-calculated optical conductivity of Mn$_3$Sn, $\sigma_{xx}(\omega)$, for various cases is displayed in Fig.~\ref{effcetivemass}(a). The nonmagnetic spectrum differs substantially from those of the constrained in-plane ferromagnetic reference state and the inverse-triangular antiferromagnetic state, whereas the diagonal conductivities of the latter two states are similar. This comparison shows only that their momentum-averaged longitudinal optical responses are similar. It neither demonstrates that the ferromagnetic reference state is realized by the applied field nor establishes the Berry-curvature origin of the anomalous Hall or Kerr response, which is encoded in the off-diagonal conductivity and must be evaluated separately.

%-------------------------------------------------------------------
%Fig3

\begin{figure*}
    \centering
    \includegraphics[width=1.0\textwidth]{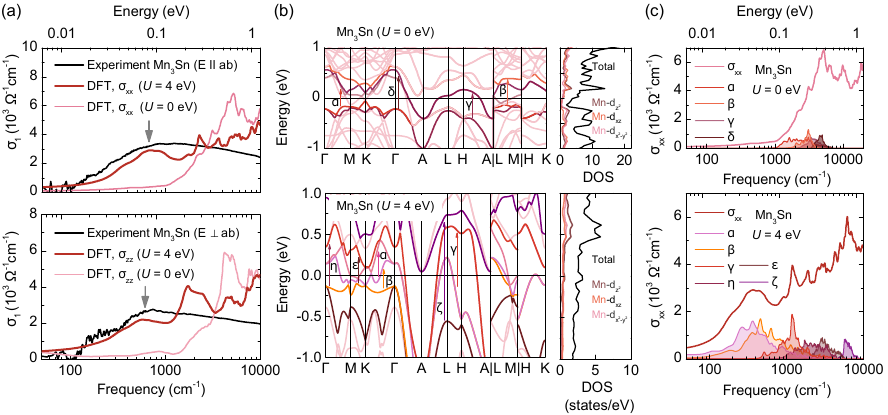}
    \caption{(a) The interband optical conductivity of Mn$_{3}$Sn for the in-plane (top) and out-of-plane (bottom) directions. The experimental conductivity ($\sigma_1$) is plotted along with the calculated optical conductivity ($\sigma_{xx}$, $\sigma_{zz}$) for Hubbard Coulomb repulsion $U = 0$ and $U = 4$~eV. The arrows indicate the appearance of low-energy interband transition for $U = 4$~eV. (b) The electronic band structure of Mn$_{3}$Sn for $U = 0$ (top) and $U = 4$~eV (bottom). The labeled arrows indicate the dominant interband transitions. (c) The band-resolved optical conductivity calculated for $U = 0$ (top) and $U = 4$~eV (bottom). The labels indicate the band-resolved conductivities whose contributions arise from the interband transitions labeled in (b).}
  \label{bandstructure}
\end{figure*}

%-------------------------------------------------------------------

Moreover, the interband spectral weight across the helical spin-ordering transition of Mn$_3$Sn is unchanged [Fig.~\ref{opticalcond}(c)]. Combined with the close similarity of the ferromagnetic and noncollinear antiferromagnetic $\sigma_{xx}(\omega)$ from our DFT calculations [Fig.~2(a)], this shows that the helical order does not significantly modify the momentum-averaged interband transitions. This transition is instead reported itself to manifest in the Berry-curvature-derived anomalous channel, in which both the anomalous Hall and magneto-optical Kerr responses are observed to collapse below the helical ordering temperature~\cite{balk2019comparing}, and it may additionally produce a minor free-carrier renormalization at THz frequencies~\cite{cheng2019terahertz}.

We have also explored the magnetic-field-dependent linear-optical response $\sigma_{xx}(\omega)$ of Mn$_3$Sn in fields
as high as $17$~T, covering the energy range from 50 to 7000~cm$^{-1}$.
We could not identify any field variation of the optical absorption [see SM~\cite{SM}]. While clean Weyl and Dirac semimetals with isolated, low-density cones are expected to display a $\sqrt{B}$ Landau-level ladder in the absorption spectrum~\cite{akrap2026landau,pronin2023linear}, the absence of any field response hints at strong free-carrier screening and correlations from non-trivial bands, with this field independence persisting even below 50~cm$^{-1}$, as observed by Cheng~\textit{et al.}~\cite{cheng2019terahertz}.

The effect of correlations on the free-carrier response can be assessed from a plasma-frequency analysis~\cite{qazilbash2009electronic, basovelectrodynamics}. The zero crossing of the real part of the dielectric function, $\varepsilon_1(\omega)$, yields the screened plasma frequency $\omega_p^{\mathrm{scr}}$, which is related to the unscreened plasma frequency via $\omega_p = \omega_p^{\mathrm{scr}}\sqrt{\varepsilon_\infty}$, where $\varepsilon_\infty$ accounts for the contribution of high-energy interband transitions, which for Mn$_3$Sn is found to be 8 $\pm$ 1. We estimated $\omega_p$ by combining the infrared data with spectroscopic ellipsometry covering frequencies up to $\sim$5~eV [see SM~\cite{SM}. The inset of Fig.~\ref{opticalcond}(c) displays $\varepsilon_1(\omega)$ for Mn$_3$Sn. We obtain plasma frequencies of approximately 3.4 and 2.8~eV (at 10~K, considering $\pm$0.3~eV error) for the in-plane and out-of-plane responses, respectively [see Table~\ref{tab:plasma}]. However, a comparison with the DFT-calculated plasma frequency, without correlation ($U = 0$) gives $(\omega_p^{\mathrm{exp}})^2/(\omega_p^{\mathrm{DFT}})^2$ $= 4.63$ and 3.18, respectively. Because a renormalization of this size ($\gg$ 1) cannot plausibly be assigned to many-body effects alone in a $3d$ intermetallic~\cite{ferber2010analysis}, the discrepancy is more naturally attributed to an inadequate description of the underlying bands, and thus motivates a finite $U$ for Mn$_3$Sn.

\newpage
\subsection{Free-carrier response of Mn$_3$Ge \\ and extended Drude analysis}

To further illuminate the effect of electronic interactions, we compare the optical properties of Mn$_3$Sn with those of isostructural Mn$_3$Ge. Owing to the smaller atomic radius of Ge, the lattice parameters of Mn$_3$Ge contract by almost 5\% relative to Mn$_3$Sn, which is expected to enhance the electron hopping. Moreover, the ordered Mn moment in Mn$_3$Ge ($\sim$2.2~$\mu_\mathrm{B}$~\cite{rai2022unconventional,soh2020ground}) is smaller than in Mn$_3$Sn ($\sim$3.0~$\mu_\mathrm{B}$~\cite{brown1990determination}), pointing to a more itinerant, less localized character in Mn$_3$Ge. Together, these structural and magnetic trends provide a clean testing ground for isolating the role of electronic correlations.

The temperature evolution of the optical conductivity of Mn$_3$Ge is shown in Fig.~\ref{opticalcond}(d).
Above $1000~\mathrm{cm}^{-1}$ the high-energy interband contributions are nearly temperature independent, similar to those of Mn$_3$Sn. However, Mn$_3$Ge has a dominant and strongly anisotropic free-carrier (Drude) contribution, based on the decomposed spectra in the inset of Fig.~\ref{opticalcond}(d). Notably, the in-plane relaxation rate at 10~K is twice that of Mn$_3$Sn. The anisotropy of the Drude response between the in-plane and out-of-plane directions is also pronounced, indicating that Mn$_3$Ge is electronically closer to a quasi-two-dimensional kagome system that exhibits a sizable optical anisotropy.

In order to get more insights into the differences of the low-energy responses of Mn$_3$Ge and Mn$_3$Sn, we performed the extended Drude analysis, which provides the frequency-dependent effective mass and scattering rate of free-carriers~\cite{dressel2002electrodynamics}. The effective mass of Mn$_3$Sn, shown in Fig.~\ref{effcetivemass}(b), reaches about $1.6\,m_b$ at 50~cm$^{-1}$, where $m_b$ denotes the band-mass. This value is close to the effective mass estimated previously ($\sim2\,m_b$)~\cite{cheng2019terahertz}. For the out-of-plane direction, the effective mass is reduced by almost half. Interestingly, the metallic region obeys a Fermi-liquid behavior, with $1/\tau \propto \omega^2$ upto at least 500~cm$^{-1}$ [see Fig.~\ref{effcetivemass}(d)], extending what was previously observed in the THz range~\cite{cheng2019terahertz}. A comparison with Mn$_3$Ge, see Fig.~\ref{effcetivemass}(c) and (e), shows that its in-plane effective mass ($\sim 0.5\,m_b$) is at most one-third of that of Mn$_3$Sn, with the same trend out of plane, while the Fermi-liquid behavior is retained. By comparison with the well known results on heavy fermions and high-$T_c$ cuprates~\cite{roy2024,degiorgi1999electrodynamic,puchkov1996pseudogap}, we conclude that Mn$_3$Sn is intermediately correlated, in contrast to the weak electronic interactions in Mn$_3$Ge.

\subsection{Correlated band structure \\ and band-resolved optical response of Mn$_3$Sn}

\begin{figure*}
    \centering
    \includegraphics[width=1.0\textwidth]{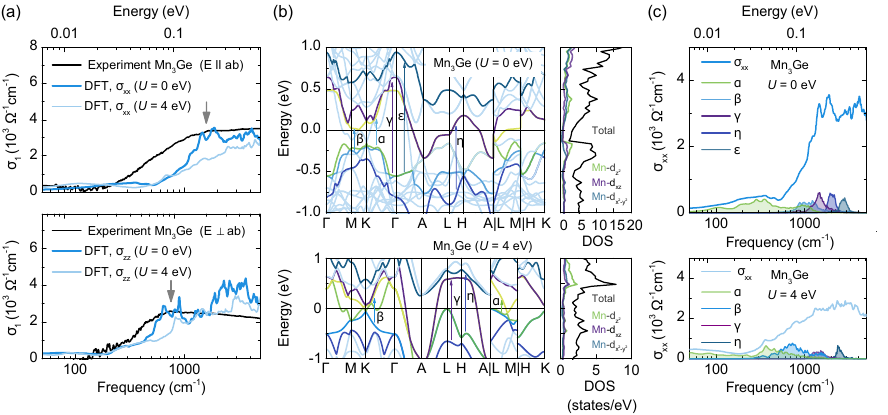}
    \caption{(a) The interband optical conductivity of Mn$_{3}$Ge for the in-plane (top) and out-of-plane (bottom) directions. The experimental conductivity ($\sigma_1$) is plotted along with the calculated optical conductivity ($\sigma_{xx}$, $\sigma_{zz}$) for Hubbard Coulomb repulsion $U = 0$ and $U = 4$~eV. The arrows indicate the anisotropic low-energy interband transitions present in the $U = 0$ case. (b) The electronic band structure of Mn$_{3}$Ge for $U = 0$ (top) and $U = 4$~eV (bottom). The labeled arrows indicate the dominant interband transitions. (c) The band-resolved optical conductivity calculated for $U = 0$ (top) and $U = 4$~eV (bottom). The labels indicate the band-resolved conductivities whose contributions arise from the interband transitions labeled in (b).}
  \label{Mn3Gebandstructure}
\end{figure*}

Based on the optical mass enhancement observed in Mn$_3$Sn, we now examine how the electron correlations reshape the band structure and the associated optical transitions. Here we apply a Hubbard $U$ to the Mn $d$-electrons, together with a Hund's coupling $J$, to model the correlations within the DFT+$U$ framework. This approach has proven to capture the electronic structure of many weakly to intermediately correlated systems, including kagome magnets~\cite{xu2020electronic,schilberth2022magneto,wenzel2022effect, sharma2025exploring}.

The electronic band structure of Mn$_3$Sn without correlations ($U = 0$) is shown in Fig.~\ref{bandstructure}(b), which exhibits multiple metallic bands crossing the Fermi level ($E_F$) along with band crossings predicted to host Weyl nodes near $E_F$, along the M--K directions. Moreover, the density of states is dominated by Mn $d$-orbital contributions, as reported previously~\cite{kubler2017weyl}.

The corresponding optical conductivity, plotted alongside the experimental interband contribution in Fig.~\ref{bandstructure}(a) (both at 10~K), shows that the low-energy interband response is absent in bare DFT calculations. However, when a Coulomb repulsion of $U = 4$~eV and a Hund's coupling of $J = 0.25$~eV are included, the low-energy optical conductivity is well reproduced. This scenario also holds for the out-of-plane direction, where the low-energy transitions (indicated by the arrows) of $\sigma_{xx}$ and $\sigma_{zz}$ are close in energy, giving a near-isotropic response, consistent with three-dimensional electronic character.

Within the parameter sets considered, the Lichtenstein DFT$+U$ calculation with $U = 4$~eV and $J = 0.25$~eV provides the closest phenomenological description of the measured low-energy interband spectrum. These values are specific to the adopted correlated subspace and projector convention and should not be interpreted as a direct determination of the bare atomic interactions or compared one-to-one with DFT+DMFT parameters.

The recent study by Cao, Xu, and Yang \cite{cao2026correlated} used full-potential WIEN2k combined with DFT+DMFT and found a strong sensitivity of the magnetic and topological band structures to Hund's coupling. Their parameters and many-body treatment are different from ours, so this sensitivity supports the need for a correlation-aware description but does not determine the $U$ and $J$ values in our static PAW-based DFT$+U$ calculation.

The origin of the interband transitions seen in Fig.~\ref{bandstructure}(a) is revealed by the band-resolved optical conductivity shown in Fig.~\ref{bandstructure}(c). Comparing this with the electronic band structure [Fig.~\ref{bandstructure}(b)], we note that the low-energy interband transitions occur between bands near $E_F$ (within $\pm 0.5$~eV). Including on-site correlations ($U = 4$~eV) does not simply shift the bands rigidly: the occupied Mn~$3d$ states are pushed down closer to $-0.5$~eV while several previously flat bands acquire pronounced dispersion and cross $E_F$, most visibly along $\Gamma$–A and A–L–H. At the same time, new band extrema develop just above $E_F$ near M, K, and H, opening additional low-energy transition channels.

The transitions labeled $\alpha$ and $\beta$ in the band structure [Fig.~\ref{bandstructure}(b) ($U = 4$~eV)] contribute primarily to this low-energy optical response. Notably, the nearly flat band, located around $- 0.17$~eV below $E_F$, contributes to the spectra up to 0.3~eV (transitions~$\beta$, $\varepsilon$, $\eta$). The high-energy interband excitations, such as those labeled $\gamma$ and $\zeta$, also remain optically accessible. When our calculations include correlations, we further observe fewer bands below the Fermi level and a broader DOS [Fig.~\ref{bandstructure}(b)], with indistinguishable in-plane ($d_{x^2-y^2}$) and out-of-plane ($d_{z^2}$, $d_{xz}$) Mn orbital contributions indicating a nearly isotropic electron density.

Moreover, the orbital character of the bands involved [Fig.~\ref{bandstructure}(b)] suggests that the optical transitions [Fig.~\ref{bandstructure}(c)] are not uniformly distributed across the Mn-$3d$ states. In an idealized trigonal-orbital basis, the Mn-$3d$ states may be grouped into the $a_1$ ($d_{z^2}$), $e_1$ ($d_{xz},d_{yz}$), and $e_2$ ($d_{xy},d_{x^2-y^2}$) sectors. The low-energy $\alpha$ feature can be associated with the splitting of two $e_1$-derived bands near M. More notably, the $\gamma$ and $\eta$ peaks coincide with transitions from $e_1$-derived occupied states to predominantly $e_2/a_1$-derived unoccupied states near K and M, respectively. The higher-energy $\zeta$ contribution arises from transitions between two strongly hybridized bands with appreciable $d_{z^2}$ character and may be viewed as a bonding–antibonding-like separation rather than a bare atomic crystal-field splitting.

It is worth to note that in the uncorrelated band structure of Mn$_3$Sn band crossings appears in the momentum region (along the M–K directions, close to K) at 40~meV, where earlier calculations predicted Weyl nodes. Upon including U, the corresponding feature shifts toward the Fermi level and becomes gapped; it should therefore be described as an avoided crossing rather than a Weyl node. Moreover, the presence of a nearly flat band, fewer bands, and a broader density of states below the Fermi level in the DFT+$U$ calculation is also qualitatively consistent with the ARPES spectra, which show broad, poorly resolved features~\cite{kuroda2017evidence}. The optical response of the correlated bands, which resembles the experimental spectra, indicates that the ground state of Mn$_3$Sn is correlated, with $U = 4$~eV and $J = 0.25$~eV, in the DFT+$U$ framework.

Finally, although Weyl nodes with linear dispersion near $E_F$ can in principle produce a linear optical conductivity~\cite{pronin2021nodal}, our band-resolved analysis shows that the interband contribution below $\sim$700~cm$^{-1}$ [Fig.~\ref{bandstructure}(a)] is instead a cumulative effect of multiple overlapping band-to-band transitions. The apparent linear regions, at any energy window in Mn$_3$Sn, therefore do not originate from direct transitions across bands with Weyl-like avoided crossings, revising the interpretation of Cao \textit{et al.}~\cite{Cao2023optical}.
This points to the necessity of band-resolved analysis for optical studies of related correlated topological systems, such as those discussed in Refs.~\cite{lai2018weyl,guo2018evidence,xu2017heavy}.

\subsection{Weakly correlated Mn$_3$Ge \\ and the plasma-frequency comparison}

We now examine the applicability of the correlated approach to Mn$_3$Ge. The $U = 0$ band structure of Mn$_3$Ge is displayed in Fig.~\ref{Mn3Gebandstructure}(b). Near the Fermi level, the DOS has dense contributions centered around $- 0.25$~eV, with the $d_{z^2}$ orbital dominating over the weaker $d_{x^2-y^2}$ contribution. Adding $U = 4$~eV, however, suppresses the sharp features in DOS below $E_\mathrm{F}$ and averages out the anisotropy of the orbital contributions.

A comparison of the calculated optical conductivity with the experimental conductivity is plotted  in Fig.~\ref{Mn3Gebandstructure}(a). The introduction of $U = 4$~eV and $J = 0.25$~eV  smears the interband optical conductivity, making it less compatible with the experimental spectrum. By contrast, $U = J = 0$ preserves the low-energy spectral weight seen in the experiment. Moreover, the peak positions of the in-plane and out-of-plane conductivity (indicated by the arrows) differ by around 700~cm$^{-1}$, evidencing a sizable optical anisotropy in Mn$_3$Ge.

The band-resolved conductivity for $U = 0$, shown in Fig.~\ref{Mn3Gebandstructure}(c), indicates that the low-energy response is dominated by the transitions labeled $\beta$, $\gamma$, and $\eta$ in the band structure, whereas the lowest-energy transition $\alpha$ contributes negligibly to the experimental conductivity. Furthermore, applying $U = 4$~eV introduces smearing of the spectral weight associated with the interband transitions, making the calculated spectrum deviate from the experimental conductivity. Hence, $U = 0$ yields a closer match to the optical response of Mn$_3$Ge, indicating a weakly correlated band structure as the natural state of the system. This result, $U = 0$, is consistent with the DFT and ARPES comparison of Changdar \textit{et al.}~\cite{changdar2024weak}.

However, the $U = 0$ description does not imply that Mn$_3$Ge is uncorrelated. The calculated conductivities ($\sigma_{xx}$, $\sigma_{zz}$) had to be scaled down by a factor of 2.5 (for both $U = 0$ and $U = 4$~eV) to closely resemble the experimental spectra. This scaling is also an indication of the presence of a weak correlation in Mn$_3$Ge. While the effect of $U$ on the optical conductivity resembles a continuous broadening of the interband transitions, the band structure shows that the Weyl-like avoided crossings along the M–K directions move farther from $E_\mathrm{F}$, from $\sim$5~meV to 40~meV when correlation is included. This is opposite to what was observed for Mn$_3$Sn.

Finally, we conclude our discussion of correlation effects by comparing the theoretical and experimental plasma frequencies, see Table~\ref{tab:plasma}. For Mn$_3$Sn, the in-plane ratio $(\omega_p^{\mathrm{exp}})^2/(\omega_p^{\mathrm{DFT+\it{U}}})^2$ is $0.80$ once correlations are included ($U = 4$~eV), compared with $4.63$ for bare DFT ($U = 0$); this large deviation from unity reflects a severe underestimate of the plasma frequency by bare DFT and again indicates that the bare DFT description is inadequate for Mn$_3$Sn, whereas the $U = 4$~eV calculation captures the experimental observations. By contrast, Mn$_3$Ge already shows close agreement between theory and experiment at $U = 0$, with an in-plane ratio of $0.72$, whereas including correlations drives the ratio well below unity, to $0.30$. This indicates that Mn$_3$Ge exhibits only weak electronic correlations, which are well described by $U = 0$.

\section{CONCLUSIONS}

Combining broadband optical spectroscopy with density functional theory, we have compared the low-energy electrodynamics of the isostructural kagome antiferromagnets Mn$_3$Sn and Mn$_3$Ge. Both compounds are good metals whose far-infrared response follows a Fermi-liquid-like behavior, $1/\tau \propto \omega^{2}$, at least up to $\sim$500~cm$^{-1}$. They differ, however, in one important respect: The Drude scattering rate of Mn$_3$Sn is nearly the same in-plane and out-of-plane, and so are its interband features, whereas both are markedly directional dependent in Mn$_3$Ge. Mn$_3$Sn therefore behaves as an electronically three-dimensional kagome metal, while Mn$_3$Ge lies closer to a quasi-two-dimensional one. The extended-Drude analysis further gives an optical effective mass of $1.6\,m_b$ at 50~cm$^{-1}$ in Mn$_3$Sn, about three times the value obtained for Mn$_3$Ge from the identical analysis.

This contrast is reproduced by the calculations only when on-site correlations are treated explicitly. A Lichtenstein DFT+$U$ description with $U = 4$~eV and $J = 0.25$~eV accounts for the measured low-energy optical response of Mn$_3$Sn, whereas Mn$_3$Ge is best described at $U = 0$. The Drude-weight analysis leads independently to the same conclusion: for Mn$_3$Sn the calculated plasma frequency rises from 1.58~eV at $U = 0$ to 3.80~eV at $U = 4$~eV, in accord with the measured value, while for Mn$_3$Ge the agreement is already good without a Hubbard correction. Correlations also reshape the band crossings that earlier work identified as Weyl nodes. In Mn$_3$Sn, adding $U$ moves the crossings toward the Fermi level and opens a gap, so that they are better described as avoided crossings, whereas in Mn$_3$Ge the same treatment pushes them away from the Fermi level. Our band-resolved decomposition shows that the apparent linear segments of $\sigma_1(\omega)$ in both compounds arise from several overlapping interband transitions rather than from direct excitations within a single Weyl-like band.

The calculations further separate the presence of the Mn moments from their arrangement. The optical conductivities obtained for the ferromagnetic and the noncollinear antiferromagnetic configurations are nearly identical, and both differ clearly from the nonmagnetic case, so the longitudinal response is governed by the ordered moments themselves rather than by how they are oriented.
This similarity does not imply that an applied field converts the noncollinear order into the ferromagnetic state, nor does it establish the Berry-curvature origin of the anomalous Hall and Kerr responses, which are carried by the off-diagonal conductivity and must be evaluated separately. Consistently, the measured infrared magneto-reflectivity of Mn$_3$Sn and Mn$_3$Ge shows no field dependence up to 17~T, as expected when strong free-carrier screening dominates the low-energy response.

Taken together, these results show that correlation strength and electronic dimensionality can differ appreciably between two compounds of the same crystal structure, and that optical spectroscopy places both parameters on a quantitative footing. Tuning them within the Mn$_3X$ family therefore offers a practical route to engineering the optoelectronic and spintronic response of related correlated materials~\cite{wen2025enhanced,zhao20263d,panda2026efficient}. They also hint that electronic correlations may be essential to the interpretation and calculation of altermagnetism and related phenomena in Mn$_3$Sn and other noncollinear compensated magnets~\cite{jungwirth2026symmetry,cheong2024altermagnetism,park2026intra,sachin2026altermagnetic}.

\vspace{1.5\baselineskip}
\section*{Author Contributions}

R.M.R. performed the optical measurements, analyzed the results, and wrote the manuscript. B.T. and X.F. performed the first-principles calculations. X.F. analyzed the computational results and prepared the corresponding theoretical figures. M.W. contributed to the optical measurements and to the discussion of the results. C.S. and C.F. provided the samples. A.T. performed the ellipsometry measurements. M.O. performed the high-field infrared measurements. A.V.P. and M.D. supervised the project. All authors discussed the results and commented on the manuscript.

\section*{Data Availability}

The data that support the findings of this article are not publicly available. The data are available from the authors upon reasonable request.

%-------------------------------------------------------------------
\begin{acknowledgments}

The authors acknowledge the technical support from Gabriele Untereiner (Universität Stuttgart) and thank Vignesh Sundaramurthy (Max Planck Institute for Solid State Research, Stuttgart, Germany) for the Laue x-ray diffraction measurements. A portion of this work was performed at the National High Magnetic Field Laboratory, which is supported by National Science Foundation Cooperative Agreement No. DMR-2128556 and the State of Florida.

\end{acknowledgments}

\newpage
\bibliography{Mn3SnGe}

\input{SM_arxiv}

\end{document}

%% file: SM_arxiv.tex
% =====================================================================
%  SM_arxiv.tex -- Supplemental Material, appended to main.tex for arXiv
%  Insert in main.tex immediately after \bibliography{Mn3SnGe}:
%        \input{SM_arxiv}
%  The SM has its OWN reference list [1], [2], ... printed at the end of
%  the SM. SM references use \smcite / \smciterange (NOT \cite), so they
%  never enter main.aux/main.bbl and do not change the main reference list.
% =====================================================================

\clearpage
\onecolumngrid

% ---- S-numbering for sections, equations, figures, tables -----------
\setcounter{section}{0}
\setcounter{equation}{0}
\setcounter{figure}{0}
\setcounter{table}{0}
\renewcommand{\thesection}{S\arabic{section}}
\renewcommand{\thesubsection}{\thesection.\arabic{subsection}}
\renewcommand{\theequation}{S\arabic{equation}}
\renewcommand{\thefigure}{S\arabic{figure}}
\renewcommand{\thetable}{S\arabic{table}}
% unique hyperref anchors (avoids "destination with the same identifier")
\renewcommand{\theHsection}{SM.\arabic{section}}
\renewcommand{\theHequation}{SM.\arabic{equation}}
\renewcommand{\theHfigure}{SM.\arabic{figure}}
\renewcommand{\theHtable}{SM.\arabic{table}}

% ---- Separate SM bibliography (independent of BibTeX/natbib) ---------
\newcounter{smref}
\renewcommand{\thesmref}{\arabic{smref}}
\newcommand{\smcite}[1]{[\ref{smref:#1}]}
\newcommand{\smciterange}[2]{[\ref{smref:#1}--\ref{smref:#2}]}
\newcommand{\smbibitem}[1]{\item\label{smref:#1}}
\newenvironment{smbibliography}{%
  \par\vspace{12pt}%
  \begin{center}\rule{0.5\textwidth}{0.5pt}\end{center}%
  \vspace{2pt}\small
  \begin{list}{[\thesmref]}{%
    \usecounter{smref}%
    \setlength{\labelwidth}{2.2em}%
    \setlength{\labelsep}{0.4em}%
    \setlength{\leftmargin}{2.6em}%
    \setlength{\itemsep}{1pt}%
    \setlength{\parsep}{0pt}%
    \setlength{\topsep}{0pt}}}%
  {\end{list}}

% ---- SM title block (\maketitle cannot be called twice) -------------
\begin{center}
{\large\bfseries Supplemental Material for:\\[2pt]
``Electronic correlations shape the low-energy optical response of the\\
kagome antiferromagnets Mn$_3$Sn and Mn$_3$Ge''}\\[8pt]
R.~Mathew Roy, Bo Tai, Maxim Wenzel, Achyut Tiwari, Mykhaylo Ozerov,
Chandra Shekhar,\\ Claudia Felser, Artem V. Pronin, Xiaolong Feng,
and Martin Dressel
\end{center}
\vspace{6pt}

% =====================================================================
\section{Details on the experimental method}
\label{sm:sec:exp}

The measurements were performed on freshly polished samples.
We note that Mn$_3$Sn oxidizes when exposed to air within about two
weeks, and that the reflectivity spectra of the oxidized surface differ
from those reported here. The far-infrared range (80 to 650~cm$^{-1}$)
was covered with a Bruker IFS 113v Fourier-transform infrared
spectrometer, and the mid- and near-infrared ranges,
$\omega/(2\pi c) > 650$~cm$^{-1}$, with a Bruker Vertex 80v attached to
a Hyperion infrared microscope. Absolute reflectivity values were
obtained using the gold overcoating technique for the former and a gold
mirror as a reference for the latter. The complex optical conductivity,
$\sigma(\omega) = \sigma_1(\omega) + {\rm i}\sigma_2(\omega)$, was then
calculated from the measured reflectivity using the Kramers--Kronig
relations, with a Drude extrapolation below 80~cm$^{-1}$ and x-ray
scattering functions for the high-energy region~\smcite{S-tanner}.

The parameters of the low-frequency Drude extrapolation were taken from
the Drude fit to the measured spectra at 300~K, so that the absolute
scale of the extrapolation is set based on our own data. The resulting
dc conductivity ($\sigma_{0}$) was compared with the
temperature-dependent electrical resistivity reported for
Mn$_3$Sn~\smciterange{S-tomiyoshi}{S-nakatsuji}
and for
Mn$_3$Ge~\smciterange{S-nayak}{S-wuttke}.
The absolute resistivities in these reports differ by up to 10\%, which
is common for intermetallic single crystals and reflects differences in
crystal quality, sample geometry, and contact configuration; their
relative temperature dependence is nevertheless consistent, all reports
showing metallic behavior with a comparable resistivity ratio between
300 and 10~K.
We therefore used the averaged, normalized temperature dependence of
the published resistivity to constrain how the Drude parameters evolve
with temperature, while anchoring the absolute conductivity to the
measured far-infrared reflectivity.
Because this procedure modifies the spectra only below 80~cm$^{-1}$,
the features discussed in the main text are insensitive to it: the
interband transitions lie above 700~cm$^{-1}$, and the zero crossing of
$\varepsilon_1(\omega)$ that defines $\omega_p^{\mathrm{scr}}$ lies well
within the measured range.
This is also the main reason why we obtain the plasma frequency from the
zero crossing of $\varepsilon_1(\omega)$ rather than from the integrated
spectral weight of the low-energy Drude-like response, which would
depend on the extrapolation.

% =====================================================================
\section{Decomposition of optical spectra}
\label{sm:sec:dl}

The intraband and interband contributions to the total optical
conductivity were modeled by adding up the responses of itinerant
charges (Drude term) and localized excitations (Lorentz terms):
\begin{equation}
\sigma(\omega) = \sigma_{\rm Drude}(\omega)
               + \sigma_{\rm Lorentz}(\omega) \quad .
\label{sm:eq:sum}
\end{equation}
The complex optical conductivity can also be expressed via the complex
dielectric permittivity
[$\varepsilon = \varepsilon_1 + {\rm i}\varepsilon_2$]:
\begin{equation}
\sigma(\omega) = -{\rm i}\omega[\varepsilon(\omega) - 1]/4\pi \quad .
\label{sm:eq:opteqn}
\end{equation}
The Drude--Lorentz approach then becomes:
\begin{equation}
\varepsilon(\omega) = \varepsilon_\infty
 - \frac{\omega^2_{p,{\rm Drude}}}
        {\omega^2 + {\rm i}\omega/\tau_{\rm Drude}}
 + \sum_j \frac{\Omega_j^2}
        {\omega_{0,j}^2 - \omega^2 - {\rm i}\omega\gamma_j} \quad .
\label{sm:eq:dl}
\end{equation}
Here, $\omega_{p,{\rm Drude}}$ and $1/\tau_{\rm Drude}$ are the plasma
frequency and the scattering rate of the itinerant carriers,
respectively. The parameters $\omega_{0,j}$, $\gamma_j$, and $\Omega_j$
describe the resonance frequency, the width, and the strength of the
$j$th Lorentzian term, respectively.

The fits to the 10~K spectra are shown in the insets of Fig.~1 of the
main text.
We used the smallest number of Lorentzian terms required to reproduce
the measured conductivity, and their parameters were found to be nearly
temperature independent, consistent with the weak temperature
dependence of the interband features discussed in the main text. The
Drude width, $1/\tau$, is the only parameter that varies appreciably
with temperature; it is plotted in Figs.~1(e) and~1(f) of the main text
and discussed there.

% =====================================================================
\section{Extended Drude analysis}
\label{sm:sec:edm}

Because the intraband and interband contributions to $\sigma(\omega)$
are not separated by a well-defined energy scale in Mn$_3$Sn and
Mn$_3$Ge, we complement the Drude--Lorentz decomposition with an
extended Drude analysis. The frequency-dependent scattering rate and
effective mass are obtained directly from the complex optical
conductivity as~\smcite{S-dressel}
\begin{align}
\Gamma(\omega) &= \frac{1}{\tau^{*}(\omega)}
  = \frac{\omega_p^{2}}{4\pi}\,
    \frac{\sigma_1(\omega)}{|\hat{\sigma}(\omega)|^{2}} ,
  \label{sm:eq:edm_gamma}\\[4pt]
\frac{m^{*}(\omega)}{m_b} &= \frac{\omega_p^{2}}{4\pi}\,
    \frac{\sigma_2(\omega)/\omega}{|\hat{\sigma}(\omega)|^{2}} .
  \label{sm:eq:edm_mass}
\end{align}
Here $\hat{\sigma}(\omega) = \sigma_1(\omega) + {\rm i}\sigma_2(\omega)$
is the complex optical conductivity obtained from the Kramers--Kronig
analysis of the measured reflectivity, and
$|\hat{\sigma}(\omega)|^{2} = \sigma_1^{2}(\omega) + \sigma_2^{2}(\omega)$.
$\omega_p$ is the unscreened plasma frequency of the free-carrier
subsystem, $\omega_p^{2} = 4\pi n e^{2}/m_b$, with $n$ the carrier
density, $e$ the elementary charge, and $m_b$ the band mass.
It is obtained here from the zero crossing of $\varepsilon_1(\omega)$
together with the high-frequency dielectric constant
$\varepsilon_\infty$, as described in the main text.
$\Gamma(\omega) = 1/\tau^{*}(\omega)$ is the frequency-dependent
momentum-relaxation rate of the renormalized quasiparticles, and
$m^{*}(\omega)/m_b$ is the corresponding mass enhancement.
Equations~(\ref{sm:eq:edm_gamma}) and~(\ref{sm:eq:edm_mass}) are
written in Gaussian units, in which $\sigma$ carries units of s$^{-1}$;
the factor $4\pi$ is absent in SI units.

% =====================================================================
%  SM figures
% =====================================================================
\begin{figure}[h!]
\centering
\includegraphics[width=0.85\textwidth]{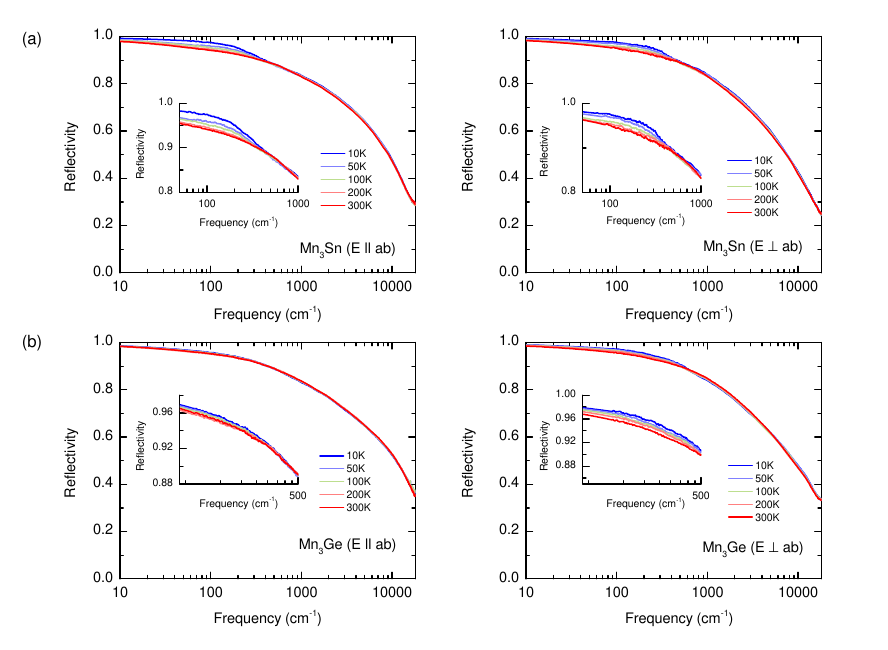}
\caption{\textbf{Temperature-dependent reflectivity:} Measured
reflectivity of (a) Mn$_3$Sn and (b) Mn$_3$Ge for the in-plane (left)
and out-of-plane (right) directions, at temperatures between 10 and
300~K. The insets show the far-infrared region on an expanded scale.
Both compounds display metallic behavior: the low-energy reflectivity
approaches unity and increases as the temperature is lowered, while the
high-frequency spectra are nearly temperature independent. The curves
below 80~cm$^{-1}$ are the Drude extrapolations used for the
Kramers--Kronig analysis, as described in Sec.~\ref{sm:sec:exp}.}
\label{sm:fig:reflectivity}
\end{figure}

\begin{figure}[h!]
\centering
\includegraphics[width=0.95\textwidth]{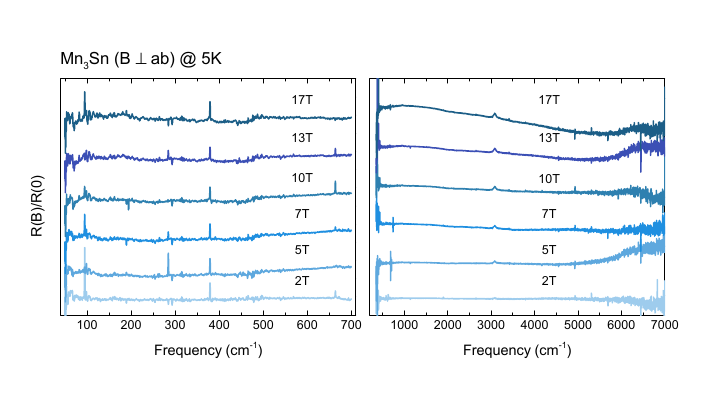}
\caption{\textbf{Magneto-optical response:} Field-induced change of the
reflectivity of Mn$_3$Sn at 5~K, plotted as the ratio $R(B)/R(0)$ for
magnetic fields between 2 and 17~T applied perpendicular to the kagome
plane ($B \perp ab$). The left and right panels cover the far-infrared
and mid-infrared ranges, respectively; the spectra are offset
vertically for clarity. No field-induced feature is resolved anywhere
between 50 and 7000~cm$^{-1}$; the sharp lines are experimental
artifacts. The same measurements were performed for $B \parallel ab$
and for Mn$_3$Ge, and none of them shows a field dependence either. The
absence of a resolvable cyclotron resonance down to 50~cm$^{-1}$ places
the cyclotron frequency below this value, consistent with a large
carrier mass and with the strong free-carrier screening expected for
the several metallic bands crossing the Fermi level, in line with the
terahertz results of Cheng \textit{et al.}~\smcite{S-cheng}.}
\label{sm:fig:field}
\end{figure}

\begin{figure}[h!]
\centering
\includegraphics[width=0.95\textwidth]{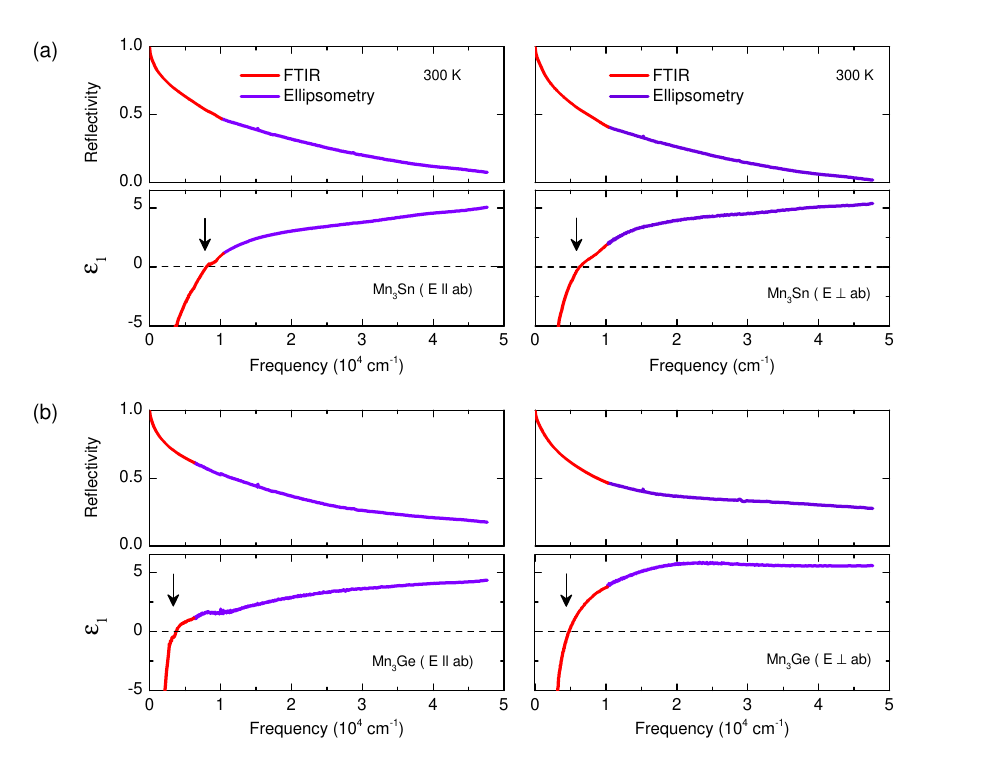}
\caption{\textbf{Merging of infrared and ellipsometric data:}
Room-temperature reflectivity (upper panels) and real part of the
dielectric function $\varepsilon_1(\omega)$ (lower panels) for
(a) Mn$_3$Sn and (b) Mn$_3$Ge, each shown for the in-plane (left) and
out-of-plane (right) directions. The infrared reflectivity measured by
FTIR spectroscopy (red) is merged with the reflectivity and dielectric
function obtained from the ellipsometric parameters (violet); the two
data sets overlap well in the region where they meet (around
10\,000~cm$^{-1}$). Extending the spectral range in this way is
required to determine the high-frequency dielectric constant
$\varepsilon_\infty$, and thereby the unscreened plasma frequency.
Arrows mark the zero crossing of $\varepsilon_1(\omega)$, which defines
the screened plasma frequency $\omega_p^{\mathrm{scr}}$.}
\label{sm:fig:ellipsometry}
\end{figure}

\begin{figure}[h!]
\centering
\includegraphics[width=0.95\textwidth]{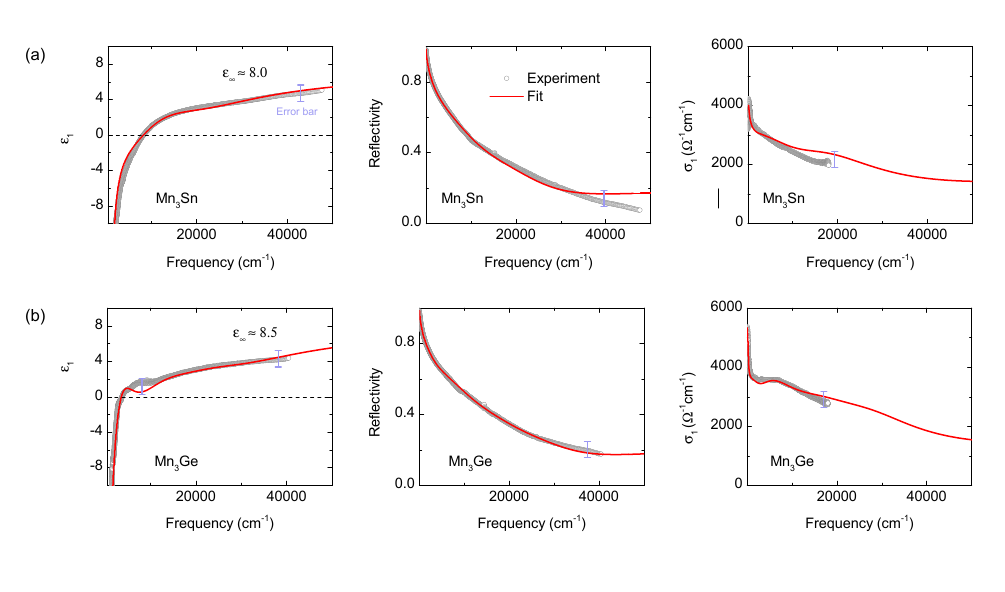}
\caption{\textbf{Determination of $\varepsilon_{\infty}$:} Simultaneous
fit of the real part of the dielectric function $\varepsilon_1(\omega)$
(left), the reflectivity (center), and the optical conductivity
$\sigma_1(\omega)$ (right) for (a) Mn$_3$Sn and (b) Mn$_3$Ge, shown at
300~K. Gray symbols are the measured data and red lines the
Drude--Lorentz fit. The fit yields $\varepsilon_{\infty} = 8.0$ for
Mn$_3$Sn and a slightly larger value of $8.5$ for Mn$_3$Ge. Since the
spectra are essentially temperature independent at these energies, the
same values are used at low temperatures. Representative error bars are
indicated; the corresponding uncertainty in the conductivity is about
10\%. Only the high-energy region is relevant for $\varepsilon_{\infty}$,
and the low-energy part of the spectra is therefore reproduced only
approximately.}
\label{sm:fig:epsinf}
\end{figure}

\begin{figure}[h!]
\centering
\includegraphics[width=0.95\textwidth]{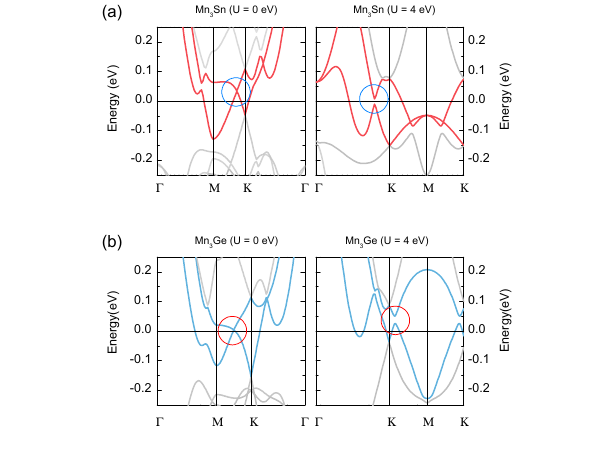}
\caption{\textbf{Weyl-like band crossings:} Close-up view of the
calculated band structures near the Fermi level, with the relevant
crossings marked by circles. (a) Mn$_3$Sn for $U = 0$~eV (left) and
$U = 4$~eV (right). Including the Hubbard correction moves the crossing
toward the Fermi level and opens a gap, so that the feature is more
appropriately described as an avoided crossing than as a Weyl node.
(b) Same for Mn$_3$Ge. Here the crossing moves away from the Fermi
level and is likewise gapped at $U = 4$~eV. The bands forming the
crossings are highlighted in color; all remaining bands are shown in
gray.}
\label{sm:fig:weyl}
\end{figure}

% =====================================================================
%  SM references (numbered separately from the main text)
% =====================================================================
\clearpage   % places all SM figures first, references after them
\begin{smbibliography}
\smbibitem{S-tanner}
D. B. Tanner, Use of x-ray scattering functions in Kramers--Kronig
analysis of reflectance,
\href{https://doi.org/10.1103/PhysRevB.91.035123}{Phys. Rev. B
\textbf{91}, 035123 (2015)}.

\smbibitem{S-tomiyoshi}
S. Tomiyoshi, H. Yoshida, T. Ohmori, and H. Yamamoto, Electrical
properties of the intermetallic compound Mn$_3$Sn,
\href{https://doi.org/10.1016/0304-8853(87)90426-4}{J. Magn. Magn.
Mater. \textbf{70}, 247 (1987)}.

\smbibitem{S-sung}
N. H. Sung, F. Ronning, J. D. Thompson, and E. D. Bauer, Magnetic phase
dependence of the anomalous Hall effect in Mn$_3$Sn single crystals,
\href{https://doi.org/10.1063/1.5021133}{Appl. Phys. Lett.
\textbf{112}, 132406 (2018)}.

\smbibitem{S-nakatsuji}
S. Nakatsuji, N. Kiyohara, and T. Higo, Large anomalous Hall effect in
a non-collinear antiferromagnet at room temperature,
\href{https://doi.org/10.1038/nature15723}{Nature \textbf{527}, 212
(2015)}.

\smbibitem{S-nayak}
A. K. Nayak, J. E. Fischer, Y. Sun, B. Yan, J. Karel, A. C. Komarek,
C. Shekhar, N. Kumar, W. Schnelle, J. K\"ubler, et al., Large anomalous
Hall effect driven by a nonvanishing Berry curvature in the noncolinear
antiferromagnet Mn$_3$Ge,
\href{https://doi.org/10.1126/sciadv.1501870}{Sci. Adv. \textbf{2},
e1501870 (2016)}.

\smbibitem{S-kiyohara}
N. Kiyohara, T. Tomita, and S. Nakatsuji, Giant anomalous Hall effect
in the chiral antiferromagnet Mn$_3$Ge,
\href{https://doi.org/10.1103/PhysRevApplied.5.064009}{Phys. Rev. Appl.
\textbf{5}, 064009 (2016)}.

\smbibitem{S-rai}
V. Rai, S. Jana, M. Meven, R. Dutta, J. Per{\ss}on, and S. Nandi,
Unconventional magnetoresistance and electronic transition in Mn$_3$Ge
Weyl semimetal,
\href{https://doi.org/10.1103/PhysRevB.106.195114}{Phys. Rev. B
\textbf{106}, 195114 (2022)}.

\smbibitem{S-wuttke}
C. Wuttke, F. Caglieris, S. Sykora, F. Scaravaggi, A. U. B. Wolter,
K. Manna, V. S\"uss, C. Shekhar, C. Felser, B. B\"uchner, et al., Berry
curvature unravelled by the anomalous Nernst effect in Mn$_3$Ge,
\href{https://doi.org/10.1103/PhysRevB.100.085111}{Phys. Rev. B
\textbf{100}, 085111 (2019)}.

\smbibitem{S-dressel}
M. Dressel and G. Gr\"uner, \textit{Electrodynamics of Solids: Optical
Properties of Electrons in Matter} (Cambridge University Press,
Cambridge, 2002).

\smbibitem{S-cheng}
B. Cheng, Y. Wang, D. Barbalas, T. Higo, S. Nakatsuji, and
N. P. Armitage, Terahertz conductivity of the magnetic Weyl semimetal
Mn$_3$Sn films,
\href{https://doi.org/10.1063/1.5093414}{Appl. Phys. Lett.
\textbf{115}, 012405 (2019)}.
\end{smbibliography}

\clearpage
% ======================= end of SM_arxiv.tex =========================